\documentclass[letter]{aa}  

\usepackage{amsmath}
\usepackage{txfonts}
\usepackage{braket}
\usepackage{amssymb}
\usepackage{enumerate}
\usepackage{graphicx}
\usepackage{subcaption}
\usepackage{graphicx,natbib}
\usepackage{txfonts}
\begin{document}

   \title{Stringent limits on the magnetic field strength in the disc of TW~Hya}

   \titlerunning{Limits to the magnetic field around TW~Hya}

   \subtitle{ALMA observations of CN polarisation}

   \author{W.~H.~T. Vlemmings
          \inst{1}\fnmsep\thanks{wouter.vlemmings@chalmers.se}
          \and B. Lankhaar \inst{1} \and P. Cazzoletti \inst{2} \and
          C. Ceccobello \inst{1} \and D. Dall'Olio \inst{1} \and
          E.~F. van Dishoeck \inst{3, 2} \and S. Facchini \inst{4}
          \and E.~M.~L. Humphreys \inst{4} \ and M.~V. Persson \inst{1} \and L. Testi \inst{4,5,6} \and J.~P. Williams \inst{7}
          }

   \institute{Department of Space, Earth and Environment, Chalmers
     University of Technology, Onsala Space Observatory, 439 92
     Onsala, Sweden
\and Max-Planck-Insitut f\"{u}r Extraterrestrische Physik,
Gie{\ss}enbachstrasse 1, 85748 Garching, Germany
\and Leiden Observatory, Leiden University, PO Box 9513, 2300 RA
Leiden, The Netherlands 
\and
European Southern Observatory, Karl-Schwarzschild-Str. 2, 85748
Garching bei M{\"u}nchen, Germany
\and
INAF – Osservatorio Astrofisico di Arcetri, Largo E. Fermi 5, 50125
Firenze, Italy
\and
Excellence Cluster Universe, Boltzmannstr. 2, 85748, Garching bei
M{\"u}nchen, Germany
\and Institute for Astronomy, University of Hawaii, Honolulu, HI
96822, USA}


   \date{13-Mar-2019}

   \abstract{ Despite their importance in the star formation process,
     measurements of magnetic field strength in proto-planetary discs
     remain rare. While linear polarisation of dust and molecular
     lines can give insight into the magnetic field structure, only
     observations of the circular polarisation produced by Zeeman
     splitting provide a direct measurement of magnetic field
     strenghts. One of the most promising probes of magnetic field
     strengths is the paramagnetic radical CN. Here we present the
     first Atacama Large Millimeter/submillimeter Array (ALMA)
     observations of the Zeeman splitting of CN in the disc of
     TW~Hya. The observations indicate an excellent polarisation
     performance of ALMA, but fail to detect significant
     polarisation. An analysis of eight individual CN hyperfine
     components as well as a stacking analysis of the strongest
     (non-blended) hyperfine components yields the most stringent limits
     obtained so far on the magnetic field strength in a proto-planetary
     disc. We find that the vertical component of the magnetic field
     $|B_z|<0.8$~mG ($1\sigma$ limit). We also provide a $1\sigma$
     toroidal field strength limit of $<30$~mG. These limits rule out
     some of the earlier accretion disc models, but remain consistent
     with the most recent detailed models with efficient advection. We
     detect marginal linear polarisation from the dust continuum, but
     the almost purely toroidal geometry of the polarisation vectors
     implies that his is due to radiatively aligned grains.}

   \keywords{Magnetic fields; Accretion, accretion disks; Stars:
     pre-main sequence; Stars: individual: TW~Hya}

   \maketitle
%

\section{Introduction}

\begin{figure*}
 \begin{minipage}[t]{0.435\textwidth}
    \includegraphics[width=\textwidth]{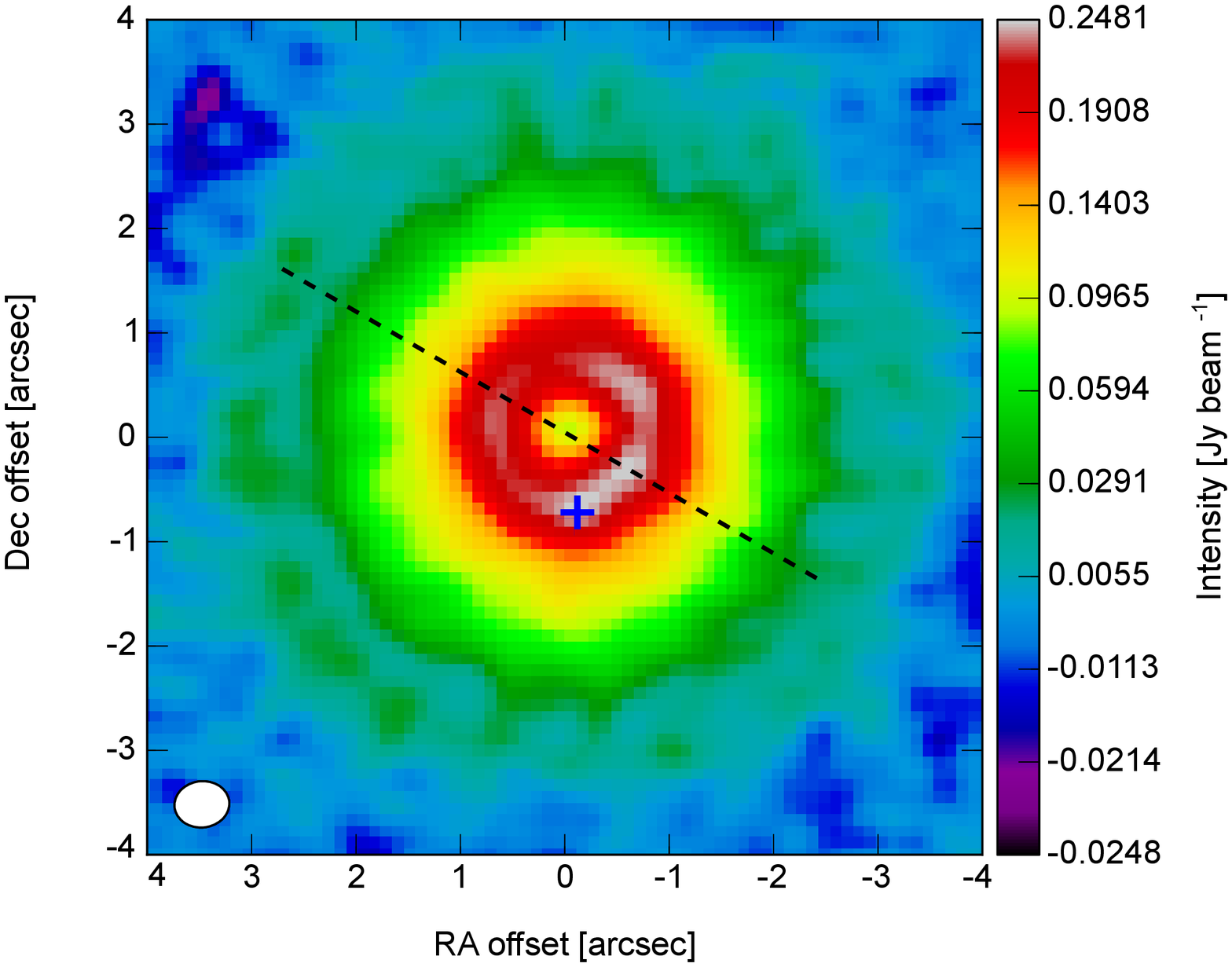}
\end{minipage}
\begin{minipage}[t]{0.435\textwidth}
    \includegraphics[width=\textwidth]{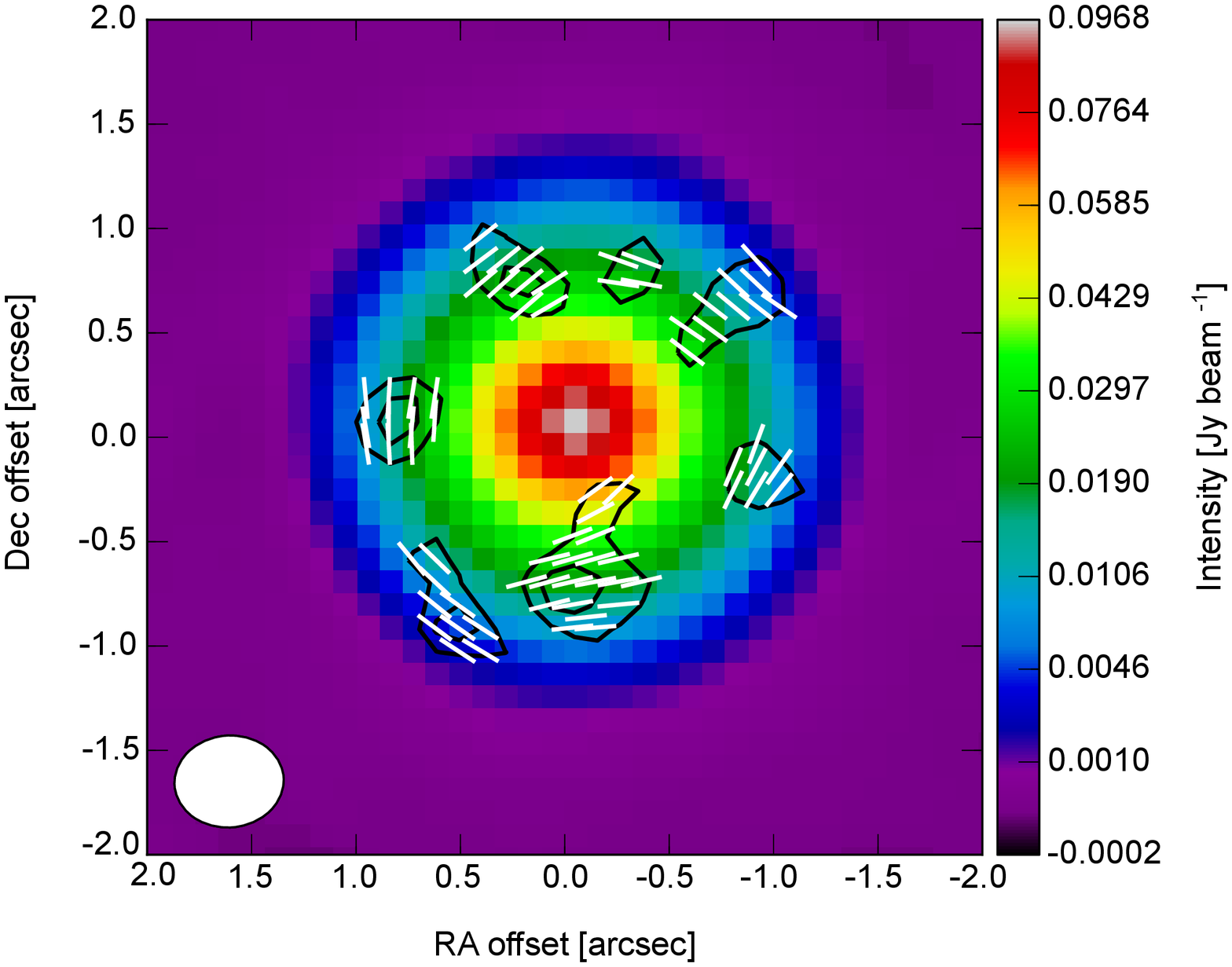}
\end{minipage}  
\hfill
\caption{({\it left:}) Integrated intensity map of the CN $(N=2-1,
 J=3/2-1/2, F=5/2-3/2)$ hyperfine transition around TW~Hya. We also
 indicate the line of nodes of the disc at a position angle of
 $240^\circ$ and the location of the strongest CN emission peak that
 was used in the polarisation analysis (see text). ({\it right:})
 Continuum and polarised 226~GHz emission of the TW~Hya disc. The
 colour scale is the total intensity dust emission. The contours,
 drawn at $3$ and $4\sigma$, are the linear polarisation, and the
 white-line segments indicate the electric vector polarisation
 angle. }
\label{map}
\end{figure*}

There have been significant efforts to observationally determine the
role of magnetic fields across all scales of star formation. For
example, through magnetic viscosity, the field plays a crucial
role in disc evolution and planet formation \citep[e.g.][]{Lizano16},
and the poloidal component of the magnetic field is responsible for
disc winds \citep{BP82}. The poloidal component of the
disc magnetic field is the result of magnetic flux that is dragged into the
disc during proto-stellar collapse, and its strength is expected to be
greater than the molecular cloud magnetic field
\citep[e.g.][]{Ferreira95, Jafari18}. However, recent models
show that the toroidal magnetic
field is likely the dominant component. Different models therefore cover a
wide range of possible values for the poloidal field, ranging from
(sub-)mG \citep[e.g.][]{Okuzumi14} to tens of mG \citep{Shu07}.

Measurements of magnetic field strengths have relied on indirect
estimates from dust polarisation and on the use of the
Chandrasekhar-Fermi method \citep[e.g.][]{Houde09} or direct
measurements of the Zeeman effect of OH, HI, or masers
\citep[e.g.][]{Vlemmings10, Crutcher12, Vlemmings17}. Additionally,
the Goldreich-Kylafis effect \citep{Goldreich1981, Goldreich1982},
observed mainly for CO, has been used to determine magnetic field
morphology \citep[e.g.][]{Cortes05}.  However, the success of magnetic
field observations of accretion or proto-planetary discs has been very
limited \citep[e.g.][]{Hughes09, Hughes13}. Furthermore, it has been
shown that self-scattering and radiative grain alignment further complicate the interpretation
of dust polarisation \citep[e.g.][]{Kataoka2015, Kataoka17}. With the exception
of a $\sim1$~kG magnetic field detection at the innermost edge (at
$\sim0.05$~au) of the disc around FU Orionis \citep{Donati05}, there
are no direct measurements of the magnetic field strength in an
accretion disc. 

One of the best probes of disc magnetic fields is the
CN radical, which is very sensitive to the Zeeman effect. Here we
present the first ALMA CN circular polarisation observations of the CN
emission arising in the disc of the T~Tauri star TW~Hya.

\subsection{TW Hya}
TW~Hya, at $d = 60$~pc \citep{Gaia18}, is a star of type K7 with an
age of $\sim8-10$~Myr. It still actively accretes, with an
accretion rate of $\sim10^{-9}$~M$_\odot$~yr$^{-1}$ \citep{Debes13}. The
star has a surface magnetic field strength of $\sim3$~kG
\citep{Donati11, Sokal18}. Its proto-planetary disc is massive, $M_{\rm
  gas}\sim0.01$~M$_\odot$, and large, extending to $\sim230$~au in the
gas lines. The TW~Hya disc has an almost face-on geometry, with
$i\sim5^\circ$ \citep{Huang18}, and the molecular lines therefore suffer
only very little line-broadening due to Keplerian motion, which results in
line widths of $\sim0.3$~km~s$^{-1}$ \citep{Teague16}. The structure of
the disc is well described \citep[e.g.][]{Bergin13, Andrews16, Kama16,
  Teague18}. Its CN emission shows a well-resolved ring-like structure
because its chemistry is driven by UV radiation \citep{Cazzoletti18}.

\subsection{CN Zeeman splitting}

The CN radical was one of the first molecules that was deteced in space. Because CN is a paramagnetic molecule, it exhibits a strong Zeeman effect
under the influence of a magnetic field. It also has a large number of
hyperfine components. We present a calculation of the exact Zeeman
splitting for the CN hyperfine components in the ALMA band 3, 6, and 7
frequency range in Appendix~\ref{CNpol}. So far, CN has been used to
measure the magnetic field in a handful of molecular clouds
\citep[e.g.][]{crutcher:96, crutcher:99, falgarone08, Hezareh10} and around a
small number of evolved stars \citep{Duthu17}.

\section{Observations and data reduction}

\begin{figure*}
 \begin{minipage}[t]{0.33\textwidth}
    \includegraphics[width=\textwidth]{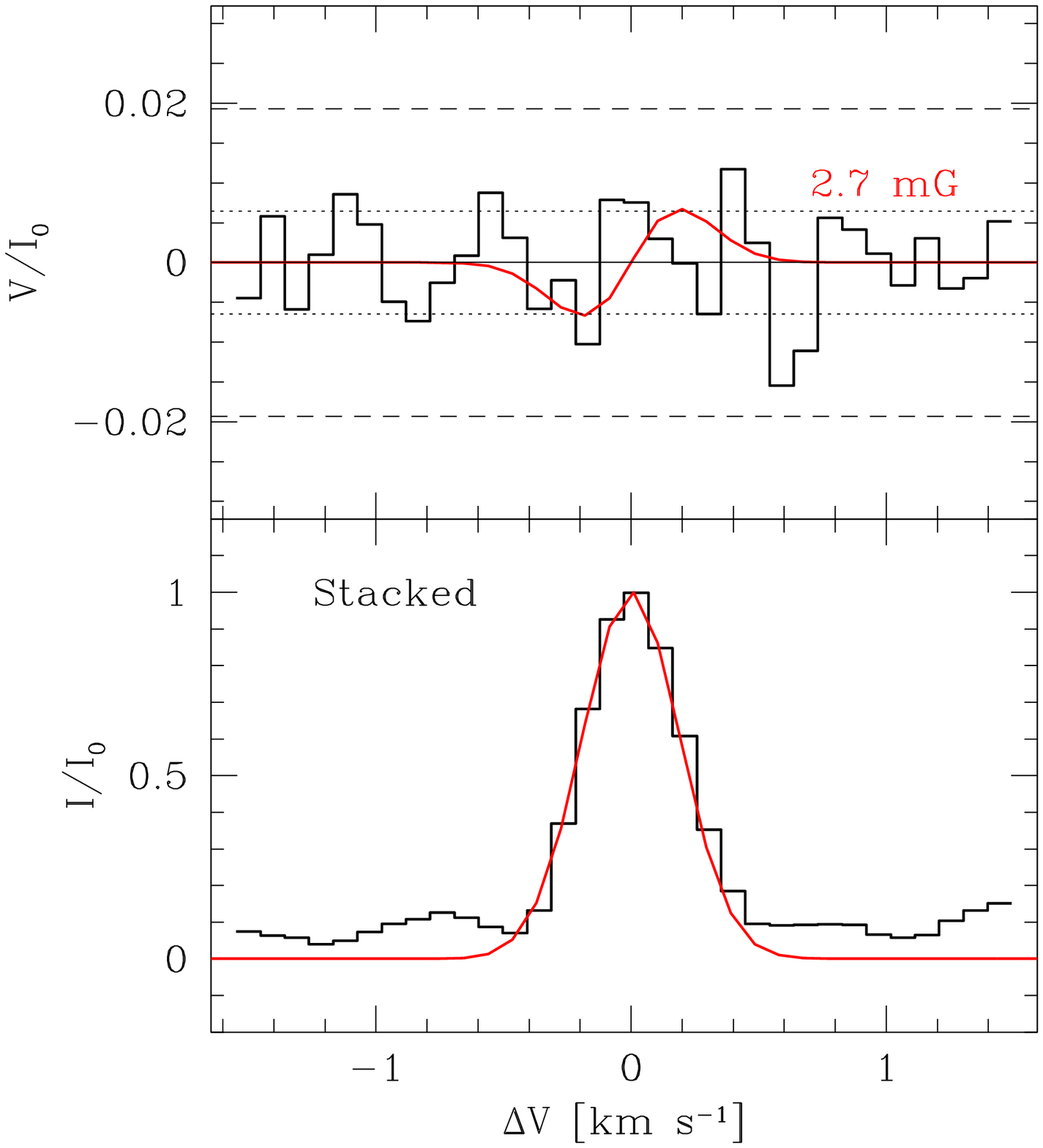}
\end{minipage}
\begin{minipage}[t]{0.33\textwidth}
   \includegraphics[width=\textwidth]{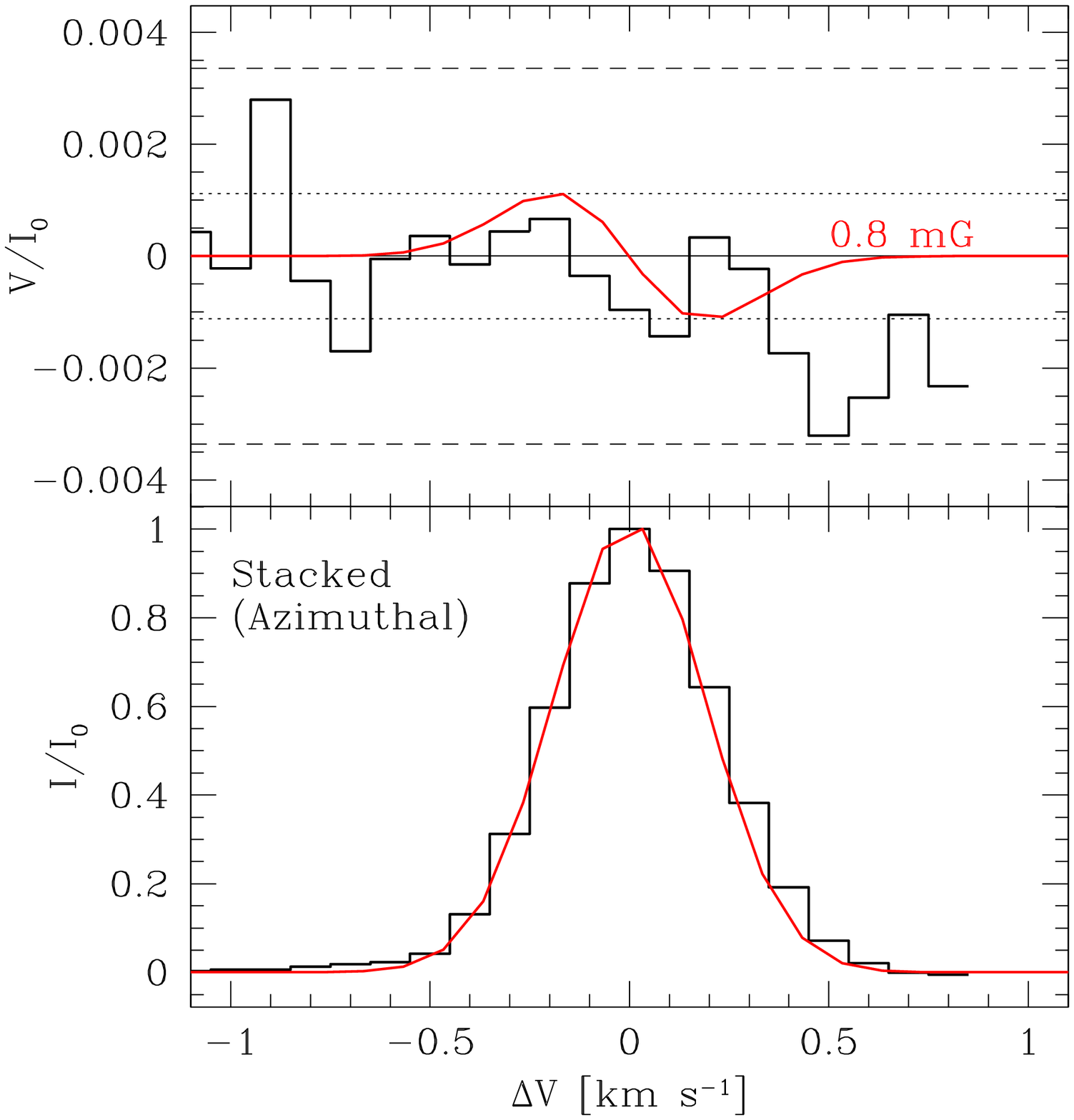}
\end{minipage}  
\begin{minipage}[t]{0.33\textwidth}
   \includegraphics[width=\textwidth]{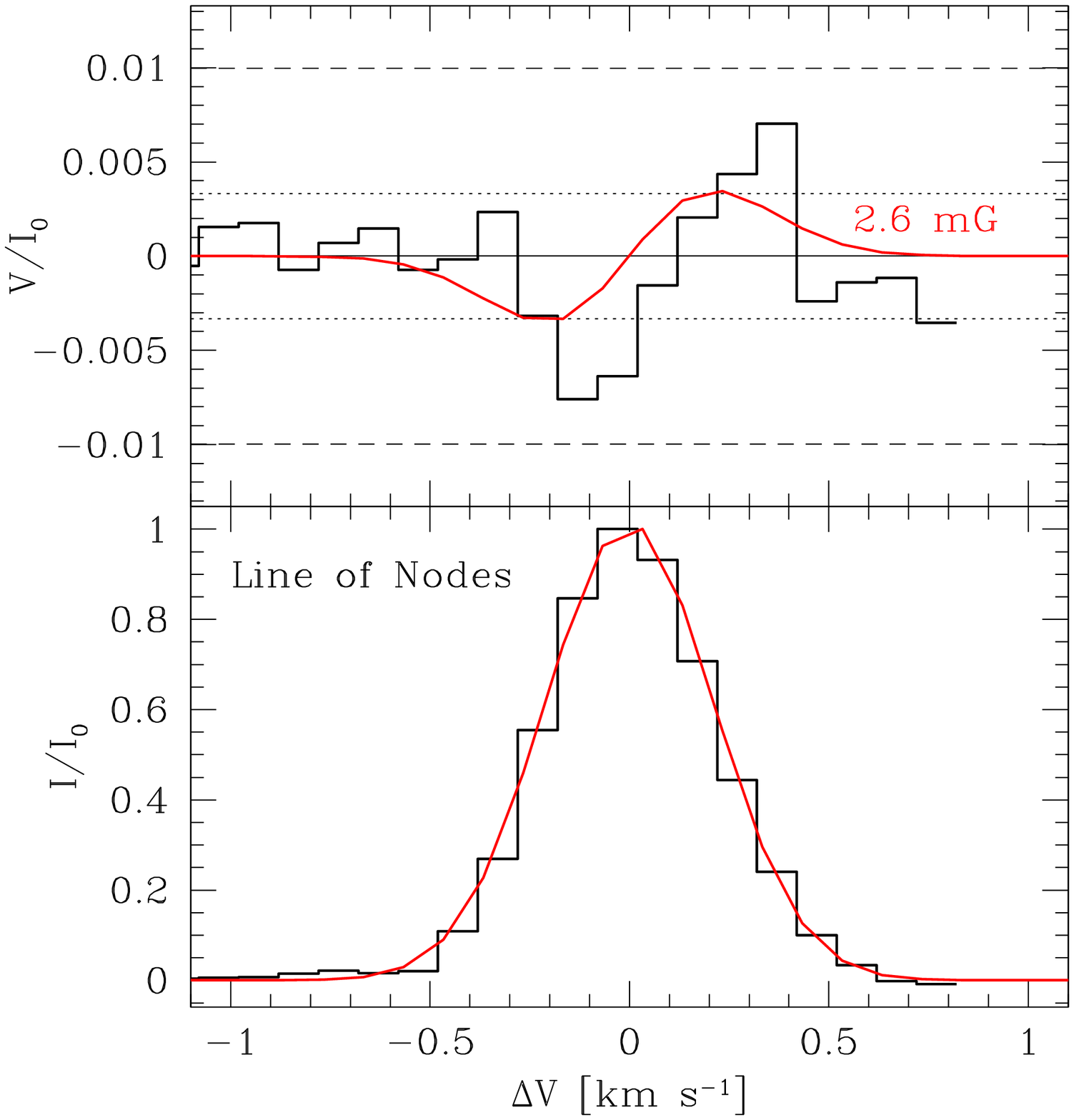}
\end{minipage}  
\hfill
\caption{Total intensity (I, bottom) and circular polarisation (V,
  top) CN spectra.  The $1\sigma$ and $3\sigma$ limits are indicated
  by the short and long dashed lines, respectively. ({\it
    left:}) Spectra stacking the nine strongest CN $N=2-1$ hyperfine
  components that we extracted at the position of the strongest emission
  peak. The $1\sigma$ limit corresponds to a magnetic field strength
  $|B_z|=2.7$~mG, for which the V-spectrum is indicated in red. The
  effective line width is $0.45~$km~s$^{-1}$. The apparent non-zero
  baseline in the Stokes I spectrum is due to the contribution to the
  stacked spectrum of hyperfine components that occur within
  $1.5$~km~s$^{-1}$ of each other. ({\it middle:}) Azimuthally
  averaged spectra obtained by stacking the two strongest
  CN hyperfine components in our data that are not significantly
  blended. The rms limit corresponds to a magnetic field strength
  $|B_z|=0.8$~mG, for which the V-spectrum is indicated in red. The
  effective line width is $0.45~$km~s$^{-1}$. Beyond $+0.5$~km~s$^{-1}$
  , the spectrum is affected by a blend with a neighbouring hyperfine
  component. ({\it right:}) Similar to the middle panel, but only for emission along the line of nodes
  (see text). Here the signature of a toroidal field would be
  strongest. The fractional rms reached on the V-spectrum is $0.33\%$.
  The rms limit corresponds to a magnetic field strength $|B|=2.6$~mG,  for which the V-spectrum is indicated in red. This corresponds to a
  toroidal field strength limit of $|B_\phi|<30$~mG. The effective
  line width is $0.5~$km~s$^{-1}$. }
\label{specs}
\end{figure*}

These observations of TW~Hya were performed as part of ALMA project
2018.1.00167.S on 2018 December 11, 12, and 13. The total observing time
in full polarisation mode was $8.8$~hr, of which $\sim3.6$~hr were
spent on TW~Hya. In order to reach the lowest possible spectral
resolution for the Zeeman splitting experiment, we tuned three
spectral windows (spws) with a width of $58.59$~MHz and $1920$
channels to cover 11 CN $(N=2-1)$ hyperfine components (see
Table.~\ref{results} for a list of observed hyperfine
components). After Hanning smoothing, this resulted in a velocity
resolution of $0.093$~km~s$^{-1}$. A fourth spw of $1.875$~GHz with
$1920$ channels and $\sim3$~km~s$^{-1}$ velocity resolution was set to
encompass the three narrow windows for optimal calibration
transfer. The initial calibration was performed using the standard
ALMA polarisation calibration scripts in CASA 5.4.0. Bandpass, flux,
and gain calibration were done using the quasars
J1107-4449 (bandpass, flux) and J1037-2934 (gain).  J1256-0547 was used for
polarisation calibration. It was noted that over the course of the
observations, the gain and polarisation calibrator showed a steady
and similar decrease in flux compared to the flux calibrator, which
was assumed to have a constant flux of 0.72~Jy at $226.64$~GHz with a
spectral index of $-0.8$. It is therefore likely that the flux calibrator
was brightening, which is also found in the ALMA calibrator
catalogue. To compensate for this change and to improve the phase
calibration at short time-intervals, we performed two rounds of phase
and one round of amplitude self-calibration using the continuum dust
emission of TW~Hya in the 1.875~GHz spectral window in which emission
lines were flagged. This improved the continuum total intensity (I)
signal-to-noise ratio (S/N) from $\sim400$ to $\sim2000$. The
self-calibration solutions were applied to all spws. Subsequently,
the continuum was subtracted using the broad spw, which was
extrapolated to the narrow spws. Finally, image cubes were created for
all spws at the native spectral resolution in all four Stokes parameters
I, Q, U, and V using Briggs weighting and a robust parameter of
$0.5$. The resulting beam size was $0.44\times0.52\arcsec$. We reach a
channel rms in the narrow spws of $\sigma=1.3$~mJy~beam$^{-1}$ for
all four Stokes parameters I, Q, U, and V. In the continuum we reach
$\sigma_{\rm (I, Q, U, V)}= (46, 22, 22, 23)~\mu$Jy~beam$^{-1}$. The
increased noise in total intensity $I$ is due to dynamic range limits.


\section{Results}
\subsection{CN polarisation}
We analysed the circular polarisation of the CN $N=2-1$ lines in a
number of different ways to extract information about the magnetic
field. Because the inclination of the disc around TW~Hya is low, we
are mostly sensitive to the vertical component of the magnetic field
$|B_z|$. A radial or toroidal magnetic field component will only
contribute $\sim10\%$ at most to $|B|$ along the minor and major
axis of the projected disc, respectively. We first analysed the
strongest CN emission peak at $(-0.09\arcsec, -0.78\arcsec)$ offset
from the continuum peak, and processed the polarisation for all the
different hyperfine components separately. The position of this peak
is indicated in the integrated intensity map of the strongest
unblended hyperfine component ($N=2-1, J=3/2-1/2, F=5/2-3/2$) shown
in Fig.~\ref{map}(left). We did not detect any significantly circular
polarisation. The limiting circular polarisation fractions with
respect to the peak total intensity emission $I_0$ and the
corresponding magnetic field limits are presented in
Table.~\ref{results}. The individual spectra are shown in
Fig.~\ref{applines}. 

Subsequently, we stacked all but the two weakest hyperfine components,
correcting for their relative intensity and sensitivity to the
magnetic field.  Specifically, the stacked Stokes V spectrum was produced
  by scaling Stokes V spectra of the individual components by the
  relative Stokes I peak intensity and the relative magnitude of the
  Zeeman coefficient with respect to that of the strongest hyperfine
  component in the stacking analysis (CN $N=2-1, J=5/2-3/2,
  F=7/2-5/2$). The resulting Stokes I and V spectra are shown in the
left panel of Fig.~\ref{specs}. In this analysis no circular
polarisation is detected either, and we reach a fractional limit ($1\sigma$)
of $|B_z|<2.7$~mG. Finally, we produced an azimuthally averaged
spectrum that was corrected on a pixel-by-pixel basis for the
Keplerian rotation of the disc. This correction is essential to
reduce the line width and optimise the detectability of the circular
polarisation. Because of the limited angular resolution, some line
broadening remains in excess of the turbulent line width derived by
\citet{Teague16}. The spectrum is taken within a ring, with a width of
the synthesised beam, that includes the brightest CN emission. The
ring has a radius of $0.72\arcsec$, which corresponds to
$\sim42$~au. The stacked spectrum is best represented using a
line width of $0.45$~km~s$^{-1}$. We limit this analysis to the two
strongest unblended hyperfine components that are most sensitive
to the magnetic field (CN $N=2-1, J=5/2-3/2,
  F=3/2-1/2$ and $N=2-1, J=3/2-1/2,
  F=5/2-3/2$). A stacked spectrum of the two lines is shown in
the middle panel of Fig.~\ref{specs}. This analysis provides the
tightest constraint on the magnetic field strength, and reaches a
$1\sigma$ limit of $|B_z|<0.8$~mG.

In order to provide a rough
estimate on the toroidal field strength, we repeated the analysis
and restricted ourselves to the CN emission in the eastern and western
part of the major axis of the projected inclined disc. This
corresponds to the line of nodes. In our analysis, we corrected for the
expected sign difference between the two sides. As shown in the right
panel of Fig.~\ref{specs}, we reach an rms of $\sim0.33\%,$ which requires a
slightly broader line width of $0.5$~km~s$^{-1}$. This corresponds to
$|B_z|<2.6$~mG or a toroidal field strength of
$|B_\phi|=|B_z|/\sin{i}<30$~mG. The stacked line of nodes of the V-spectrum
does display a $2\sigma$ signal that could correspond to a magnetic
field strength of $\sim5$~mG, although it is slightly shifted (by
$\sim0.1$~km~s$^{-1}$) from the expected location. If this is real, it should
originate from the toroidal magnetic field because no signature is seen
in the more sensitive analysis of the azimuthally averaged
spectrum. The expected reversal of a toroidal field is hinted at in
the eastern and western parts of the line of nodes separately, as shown
in Fig.~\ref{applon}, and the field strength would correspond to
$|B_\phi|\approx57$~mG. However, considering the low significance of
the circular polarisation signal ($\sim2\sigma$), we do not classify it
as a detection.

\subsection{Continuum polarisation}
Although the observation setup was optimised for a line polarisation
study, we were able to produce a linear polarisation image of the dust
continuum emission at 226~GHz using the line-free channels of the 1.875~GHz wide
spw. We debiased the linear polarisation map by calculating
the linearly polarised intensity from the Stokes Q and U images using
$P_l=\sqrt{{\rm Q}^2+{\rm U}^2 - \sigma_P^2}$.
 Here $\sigma_P$ is the
polarised rms, which can be calculated on a pixel-by-pixel basis using
$\sigma_P=\sqrt{(({\rm Q}\sigma_{\rm Q})^2 + ({\rm U}\sigma_{\rm U})^2)/({\rm Q}^2+{\rm U}^2)}.$
Here we adopt the $\sigma_{\rm Q,U}$ as measured in the images. We
find $\sigma_P\approx22~\mu$Jy~beam$^{-1}$. The
resulting polarisation map is shown in Fig.~\ref{map}(right). The
polarisation peaks at slightly above $4\sigma$ and ranges from
$<0.1\%$ of Stokes I towards the inner part of the disc to $\sim2\%$ in the outer
parts. The patchy appearance of the polarisation is likely due to the
low S/N of the observations. The polarisation vectors are almost
purely toroidal, which is the signature of radiatively aligned grains
dominating the polarised emission \citep[e.g.][]{Kataoka17}. However, because of the low
S/N of the polarisation and the lack of data at other wavelengths\footnote{The
TW~Hya disc was not detected in previous SMA continuum polarisation
observations by \citet{Hughes09}} , a discussion of the nature of the
continuum polarisation is beyond the scope of this paper.




\section{Discussion and conclusions}

\begin{figure}
\centering
\includegraphics[width=0.47\textwidth]{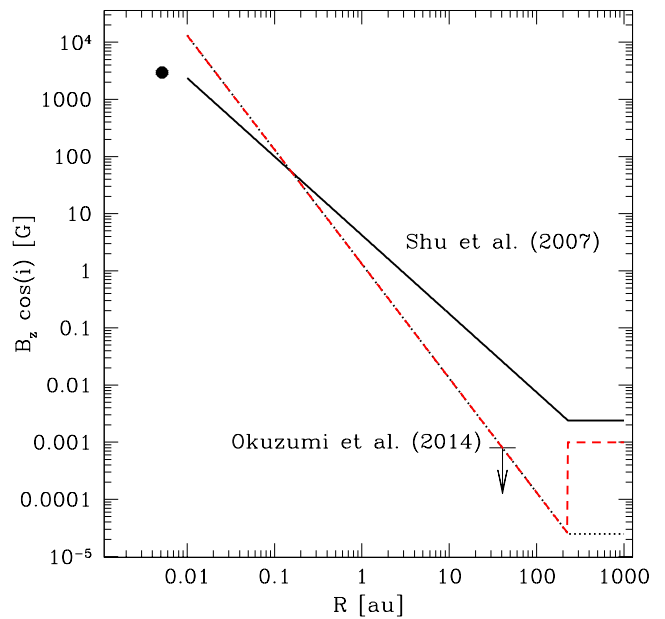}
\caption{Vertical magnetic field strength ($B_z$) from accretion
  disc models, compared with the $1\sigma$ limit obtained in our ALMA
  observations ($B_z<0.8$~mG, indicated by the arrow). The solid line
  is the T~Tauri model from Shu et al. (2007). The dotted line
  is the maximum field strength $B_{z, {\rm max}}$ from Okuzumi et
  al. (2014), taking a magnetic field strength at the outer edge of the
  disc ($R_{\rm out}=230$~au) of $B_{\infty}=25~\mu$G. The red long dashed
  line assumes $B_{\infty}=1~$mG, which corresponds to the field strength
  measured in proto-stellar cores and dense parts of molecular clouds, and it includes efficient
  advection with $D=0.025$ (see text). At $R<230$~au, the
    latter two models are identical. The solid circle shows the
  surface magnetic field measured on TW~Hya (e.g. Sokal et al. 2018).
}
 \label{Bmodelcomp}
\end{figure}

We have obtained a tight upper limit to the vertical magnetic field
component using the CN emission in the disc around TW~Hya. Our
azimuthal average field limit of $|B_z|<0.8$~mG was obtained at a
radius of approximately $42$~au. According to the model of CN emission
from \citet{Cazzoletti18}, at this radius, we are most sensitive to
the magnetic field at a disc scale height of $|z|\sim18$~au. In terms
of plasma $\beta$, the ratio of gas to magnetic pressure, this means
for the vertical field component that $\beta\gtrsim4\times10^3$ when
we adopt the disc model used in \citet{Bai15}.

The vertical magnetic field in a proto-planetary accretion disc is
generally assumed to be advected inwards from the surrounding
cloud. However, it was noted that under the influence of turbulence, a
large-scale field would diffuse away at timescales much shorter than
the advection timescales \citep[see e.g.][and references therein for a
more detailed discussion]{Okuzumi14, Guilet14}. The details of
advection and diffusion therefore strongly affect the vertical magnetic field
strength. In Fig.~\ref{Bmodelcomp} we compare our derived upper limits
with predictions from the T~Tauri star model of \citet{Shu07} and the
maximum vertical field strength according to the thin accretion disc
model from \citet{Okuzumi14}. It is immediately apparent that our
limit lies significantly below the value predicted in
\citet{Shu07}. Both models depend critically on the assumed field
strength outside the disc, $B_{\infty}$. In particular,
\citet{Okuzumi14} stated that $B_{z, {\rm max}}=(R_{\rm out}/R)^2
B_{\infty}$. Because we measure our limit at $R\sim42$~au and $R_{\rm
  out}\sim230$~au for the disc of TW~Hya, our limit would thus imply
$B_{\infty}<25~\mu$G. While such fields are consistent with Zeeman
measurements in diffuse molecular clouds \cite[e.g.][]{Crutcher12},
the fields measured in proto-stellar envelopes and dense cloud regions
are much higher and of order $1-10$~mG \citep[e.g.][]{Girart06,
  Houde09, Vlemmings10}.

Depending on the details of advection and diffusion, the vertical
field strength should be described as ${B_z=D(R)\cdot(R_{\rm out}/R)^2
B_{\infty}}$, where $D(R)$ is defined as the inverse ratio of advection
to diffusion timescales \citep[e.g.][]{Okuzumi14}. In the case of
effective advection that overcomes the diffusion of the magnetic
field, $D(R)<<1$ in the disc (where $R<R_{\rm out}$). In the case of
TW~Hya, we thus find that $D(42~{\rm au})<0.025/B_{\infty}$ for
$B_{\infty}$ given in mG. Even though this limit on $D$ depends on the
unknown magnetic field strength in the original proto-stellar envelope
around TW~Hya, our observations present the first observational
constraint of the advection in an accretion disc.

A number of studies have also related the vertical magnetic field strength
to the disc accretion rate \citep[e.g.][]{Okuzumi11, Bai13,
  Simon13}. Using the relation obtained for magneto-rotational
instability-driven accretion in the case of only Ohmic diffusion
\citep{Okuzumi11, Okuzumi14}, our limit implies
$\dot{M}\lesssim3\times10^{-7}$~M$_\odot$~yr$^{-1}$. A limit of
$\dot{M}\lesssim8\times10^{-8}$~M$_\odot$~yr$^{-1}$ is obtained when models are used in which ambipolar diffusion is taken into account
\citep{Bai13, Simon13}. These limits are fully consistent with the
accretion rates derived for the evolved TW~Hya disc.

These limits are determined using models that assume a fixed
configuration of the magnetic field without a toroidal component
and/or without treating thermodynamics. Relaxing these assumptions can
lead to a dominant toroidal field component with a strength of
$\sim10-15$~mG \citep{Bai15}, which is consistent with the toroidal
field limits we derived.

Despite the non-detection, our observations show that ALMA is able to
reach detection levels in circular polarisation of $<0.8\%$ of the
total intensity. The stacking analysis further shows no sign of
instrumental effects at even lower levels. This shows that for relatively
strong and narrow CN lines, observational limits of $<1$~mG can be
reached. 

\begin{acknowledgements}
  We thank the Nordic ARC node staff for excellent support in the
  reduction of the data. This work was supported by ERC consolidator
  grant 614264. WV and BL also acknowledge the Swedish Research
  Council (VR). SF acknowledges an ESO Fellowship. LT acknowledges
  partial support from the Italian Ministero dell\'\,Istruzione,
  Universit\`a e Ricerca through the grant Progetti Premiali 2012 --
  iALMA (CUP C52I13000140001), the Deutsche
  Forschungs-gemeinschaft (DFG, German Research Foundation) - Ref
  no. FOR 2634/1 TE 1024/1-1, and the DFG cluster of excellence
  Origin and Structure of the Universe.  This paper makes use of the
  following ALMA data: ADS/JAO.ALMA\#2018.1.00167.S. ALMA is a
  partnership of ESO (representing its member states), NSF (USA) and
  NINS (Japan), together with NRC (Canada), NSC and ASIAA (Taiwan),
  and KASI (Republic of Korea), in cooperation with the Republic of
  Chile. The Joint ALMA Observatory is operated by ESO, AUI/NRAO and
  NAOJ.
\end{acknowledgements}

\bibliographystyle{aa}

\begin{appendix}

\section{Zeeman-splitting coefficients of CN}
\label{CNpol}

The Zeeman parameters of CN have been reported for some transitions, but a comprehensive list has so far not been given. \citet{bel:89} reported Zeeman parameters for the strongest components of the $N=1-0$ and $N=2-1$ manifold, and referred to \citet{gordy:84} as their method. Referring to the same method, \citet{crutcher:96} reported slightly different Zeeman parameters for the $N=1-0$ manifold of CN. In the following, we outline the basic approach that is required in modelling the Zeeman effects of radicals. We base our methods on the theory given in \citet{brown:03} and summarise our calculations by reporting the Zeeman-splitting factors of the lines relevant to the domain of ALMA polarisation observations.

To later model the Zeeman effects of CN, it is important to first focus on its fine structure at zero magnetic field. CN has one unpaired electron, and in its ground-electronic state ($^2\Sigma$), in which the microwave lines relevant to our purpose occur, the total electron spin is $S=\frac{1}{2}$ and the electrons have no total orbital angular momentum projection $|\Lambda|=0$. The relevant interactions that introduce spectral fine structure are mediated through the molecular rotational motion, $\hat{H}_{\mathrm{rot}}$ and the spin-rotation interaction, $\hat{H}_{\mathrm{sr}}$. Moreover, the nitrogen nucleus has an intrinsic magnetic moment from its nuclear spin. Additional interactions involving the nuclear spin of the nitrogen nucleus, $\hat{H}_{\mathrm{hyp}}$, are a quadrupole interaction: nuclear spin-rotation interaction and a Fermi-contact interaction between the nuclear and electron spins. These interactions combined lead to the total effective fine-structure Hamiltonian of CN:
\begin{align}
\hat{H}_{\mathrm{eff}} = \hat{H}_{\mathrm{sr}} + \hat{H}_{\mathrm{rot}} + \hat{H}_{\mathrm{hyp}}.
\label{eq:zero_eff}
\end{align}  
In this effective Hamiltonian, the interactions can be represented by effective coupling constants belonging to the coupling of angular momentum operators for the electron and nuclear spin and the molecular rotation. Matrix elements for the Hamiltonian of Eq.~(\ref{eq:zero_eff}) are obtained in a Hund's case (b) angular momentum basis \citep{dixon:77, brown:03}. The basis functions are denoted as $\ket{\eta N S J I F M_F}$, where $N$ stands for the rotational angular momentum, $S$ for the spin-angular momentum, $J$ for the total angular momentum, $I$ for the nuclear spin angular momentum, and $F$ is the total hyperfine angular momentum. $M_F$ is the projection of the total hyperfine angular momentum on the projection axis, and all other quantum numbers are collected in $\eta$. We refer to \citet{brown:03} (Eqs.~10.52-56) for the matrix elements of Eq.~(\ref{eq:zero_eff}) in the Hund's case (b) basis. Matrix elements of Eq.~(\ref{eq:zero_eff}) in this basis are diagonal in $I$, $F,$ and $M_F$, and in very good approximation diagonal in $N$ \citep{dixon:77}. However, strong mixing occurs between the two $J=N\pm \frac{1}{2}$ states. 

We treat Eq.~(\ref{eq:zero_eff}) as being diagonal in $N$, but having mixing $J=N\pm \frac{1}{2}$ basis functions. Diagonalisation of Eq.~(\ref{eq:zero_eff}) then yields the eigenfunctions 
\begin{align}
\ket{N (J=N-\frac{1}{2}) F M_F} &= \cos \phi_{\eta NF} \ket{\eta N S (J=N-\frac{1}{2}) I F M_F} \nonumber \\
&+ \sin \phi_{\eta NF}  \ket{\eta N S (J=N+\frac{1}{2}) I F M_F} \nonumber \\
\ket{N (J=N+\frac{1}{2}) F M_F} &= \cos \phi_{\eta NF} \ket{\eta N S (J=N+\frac{1}{2}) I F M_F} \nonumber \\ 
&- \sin \phi_{\eta NF}  \ket{\eta N S (J=N-\frac{1}{2}) I F M_F},
\label{eq:eig_zero}
\end{align} 
where the diagonalisation angle $\phi_{\eta NF}$ can be computed from the ratio of the off- and diagonal elements
\begin{align}
\tan \left( 2\phi_{\eta NF} \right) = \frac{2H^{\eta NF}_{J=N-\frac{1}{2},J'=N+\frac{1}{2}}}{H^{\eta NF}_{J=N-\frac{1}{2},J'=N-\frac{1}{2}} + H^{\eta NF}_{J=N+\frac{1}{2},J'=N+\frac{1}{2}}}.
\end{align}
We have used a convenient notation for the matrix elements of the Hamiltonian \citep[see][]{larsson:19}. 

The magnetic sub-levels of the fine-structure levels of CN will split up under the influence of an external magnetic field according to the Zeeman effect. The Zeeman effect of CN has contributions from the electron spin, $\hat{H}_{\mathrm{bs}}$, the molecular rotation, $\hat{H}_{\mathrm{br}}$, and the nuclear spin of the nitrogen-nucleus, $\hat{H}_{\mathrm{bi}}$. Rotational and nuclear spin Zeeman effects are higher-order Zeeman effects and scale with the nuclear magneton ($\sim 0.47$ kHz/G), whereas the electron spin Zeeman effect scales with the Bohr magneton ($\sim 1.4$ MHz/G). The Zeeman Hamiltonian of CN is 
\begin{align}
\hat{H}_{\mathrm{Zeeman}} &= \hat{H}_{\mathrm{bs}} + \hat{H}_{\mathrm{br}} + \hat{H}_{\mathrm{bi}} \nonumber \\ 
&= \mu_B |B| \left(g_S \hat{S}_z + g_r \hat{N}_z + g_I \hat{I}_z\right),
\label{eq:zee}
\end{align}
where $g_S$, $g_r$ , and $g_I$ are the spin, rotational, and nuclear-spin
$g$-factors, the magnetic field strength is given by $|B|$, $\mu_B$ is
the Bohr magneton, and $\hat{S}_z$, $\hat{N}_z$ , and $\hat{I}_z$, are
the projection elements of the spin, rotation, and nuclear-spin
operators. The Zeeman Hamiltonians matrix elements in the case (b)
basis can be found in \citet{brown:03}. We define the $g$-factor for 
a particular state $\ket{NJF}$ from the relation \citep{larsson:19} 
\begin{align}
g_{NJF} = \frac{\braket{NJF M_F|\hat{H}_{\mathrm{Zeeman}}|NJF M_F}}{\mu_B |B| M_F}.
\end{align}
The level-specific $g$-factors are evaluated by computing the eigenfunctions of 
Eq.~(\ref{eq:eig_zero}) using 
the molecular constants of \citet{dixon:77} and subsequently evaluating 
Eq.~(\ref{eq:zee}). In our calculations, we use the $g$-factors $g_S=2.0023$ and 
$g_I=\frac{0.404}{1836}$. We assume a negligible rotational contribution to the Zeeman 
effect. The effective \textit{g}-factor for a certain transition, $\overline{g}$ can be 
obtained from the \textit{g}-factors of the upper $g_1$ and lower $g_2$ levels 
by \citep{degl:06}
\begin{align}
\overline{g} = \frac{g_1 + g_2}{2} + \frac{1}{4} (g_1 - g_2) \left( F_1(F_1+1) - F_2(F_2+1) \right),
\label{eq:ave}
\end{align}
where $F_1$ and $F_2$ are the total hyperfine angular momentum of the
upper and lower states. Equation~\ref{eq:ave} is only applicable if
the Zeeman effect is significantly smaller than the line width\footnote{For a
magnetic field of $10$~mG, the largest splitting for the $N=1-0$
manifold is of order $0.06$~km~s$^{-1}$ compared to a typical
line width $\gtrsim0.4$~km~s$^{-1}$.}. In Table \ref{tab:CN_mol} we
report Zeeman coefficients for energy levels of CN relevant to
astrophysically observed CN transitions in ALMA bands 3, 6, and 7. In
accordance with the literature (e.g.~\citet{bel:89} and
\citet{crutcher:96}), we report the Zeeman parameters in terms of
Zeeman-splitting coefficients, $z = 2 \mu_B \overline{g}$.

\begin{table*}[t]
\caption{Zeeman parameters, frequencies, and Einstein coefficients of CN transitions that are relevant to ALMA polarisation measurements.}
\label{tab:CN_mol}
\centering
\begin{tabular}{l l l l l l c c c}
\hline \hline  
$N$ & $J$ & $F$ & $N'$ & $J'$ & $F'$ & $\nu$ [GHz]$^a$ & $z$ [Hz/$\mu$G] & $A \times 10^6$ [s$^{-1}$]$^a$ \\ \hline
$ 1  $&$  1/2 $&$   1/2 $&$   0  $&$  1/2  $&$  1/2  $& $ 113.1233701 (58) $ & $-0.62 $  & $  1.29$  \\ 
$ 1  $&$  1/2 $&$   1/2 $&$   0  $&$  1/2  $&$  3/2  $& $ 113.1441573 (57) $ & $ 2.18 $  & $ 10.53$  \\ 
$ 1  $&$  1/2 $&$   3/2 $&$   0  $&$  1/2  $&$  1/2  $& $ 113.1704915 (39) $ & $-0.3  $  & $  5.14$  \\ 
$ 1  $&$  1/2 $&$   3/2 $&$   0  $&$  1/2  $&$  3/2  $& $ 113.1912787 (34) $ & $ 0.63 $  & $  6.68$  \\ 
$ 1  $&$  3/2 $&$   3/2 $&$   0  $&$  1/2  $&$  1/2  $& $ 113.4881202 (33) $ & $ 2.17 $  & $  6.74$  \\ 
$ 1  $&$  3/2 $&$   5/2 $&$   0  $&$  1/2  $&$  3/2  $& $ 113.4909702 (24) $ & $ 0.56 $  & $ 11.92$  \\ 
$ 1  $&$  3/2 $&$   1/2 $&$   0  $&$  1/2  $&$  1/2  $& $ 113.4996443 (28) $ & $ 0.62 $  & $ 10.63$  \\ 
$ 1  $&$  3/2 $&$   3/2 $&$   0  $&$  1/2  $&$  3/2  $& $ 113.5089074 (28) $ & $ 1.62 $  & $  5.19$  \\ 
$ 1  $&$  3/2 $&$   1/2 $&$   0  $&$  1/2  $&$  3/2  $& $ 113.5204315 (44) $ & $ 1.56 $  & $  1.30$  \\ 
$ 2  $&$  3/2 $&$   1/2 $&$   1  $&$  3/2  $&$  1/2  $& $ 226.2874185 (69) $ & $ 0.62 $  & $ 10.30$  \\ 
$ 2  $&$  3/2 $&$   1/2 $&$   1  $&$  3/2  $&$  3/2  $& $ 226.2989427 (68) $ & $ 2.17 $  & $  8.23$  \\ 
$ 2  $&$  3/2 $&$   3/2 $&$   1  $&$  3/2  $&$  1/2  $& $ 226.3030372 (64) $ & $-1.8  $  & $  4.17$  \\ 
$ 2  $&$  3/2 $&$   3/2 $&$   1  $&$  3/2  $&$  3/2  $& $ 226.3145400 (500) $ & $ 0.27 $  & $  9.90$  \\ 
$ 2  $&$  3/2 $&$   3/2 $&$   1  $&$  3/2  $&$  5/2  $& $ 226.3324986 (56) $ & $ 2.58 $  & $  4.56$  \\ 
$ 2  $&$  3/2 $&$   5/2 $&$   1  $&$  3/2  $&$  3/2  $& $ 226.3419298 (54) $ & $-2.2  $  & $  3.16$  \\ 
$ 2  $&$  3/2 $&$   5/2 $&$   1  $&$  3/2  $&$  5/2  $& $ 226.3598710 (500) $ & $ 0.22 $  & $ 16.08$  \\ 
$ 2  $&$  3/2 $&$   1/2 $&$   1  $&$  1/2  $&$  3/2  $& $ 226.6165714 (53) $ & $-0.3  $  & $ 10.73$  \\ 
$ 2  $&$  3/2 $&$   3/2 $&$   1  $&$  1/2  $&$  3/2  $& $ 226.6321901 (35) $ & $-0.72 $  & $ 42.59$  \\ 
$ 2  $&$  3/2 $&$   5/2 $&$   1  $&$  1/2  $&$  3/2  $& $ 226.6595584 (26) $ & $-0.71 $  & $ 94.67$  \\ 
$ 2  $&$  3/2 $&$   1/2 $&$   1  $&$  1/2  $&$  1/2  $& $ 226.6636928 (25) $ & $-0.62 $  & $ 84.65$  \\ 
$ 2  $&$  3/2 $&$   3/2 $&$   1  $&$  1/2  $&$  1/2  $& $ 226.6793114 (31) $ & $-1.18 $  & $ 52.68$  \\ 
$ 2  $&$  5/2 $&$   5/2 $&$   1  $&$  3/2  $&$  3/2  $& $ 226.8741908 (23) $ & $ 0.71 $  & $ 96.22$  \\ 
$ 2  $&$  5/2 $&$   7/2 $&$   1  $&$  3/2  $&$  5/2  $& $ 226.8747813 (30) $ & $ 0.4  $  & $114.32$  \\ 
$ 2  $&$  5/2 $&$   3/2 $&$   1  $&$  3/2  $&$  1/2  $& $ 226.8758960 (20) $ & $ 1.18 $  & $ 85.87$  \\ 
$ 2  $&$  5/2 $&$   3/2 $&$   1  $&$  3/2  $&$  3/2  $& $ 226.8874202 (29) $ & $ 1.47 $  & $ 27.31$  \\ 
$ 2  $&$  5/2 $&$   5/2 $&$   1  $&$  3/2  $&$  5/2  $& $ 226.8921280 (27) $ & $ 1.06 $  & $ 18.10$  \\ 
$ 2  $&$  5/2 $&$   3/2 $&$   1  $&$  3/2  $&$  5/2  $& $ 226.9053574 (44) $ & $ 0.79 $  & $  1.13$  \\ 
$ 2  $&$  5/2 $&$   5/2 $&$   1  $&$  1/2  $&$  3/2  $& $ 227.1918195 (49) $ & $ 2.2  $  & $0.0015$  \\ 
$ 3  $&$  5/2 $&$   3/2 $&$   2  $&$  5/2  $&$  3/2  $& $ 339.4467770 (500) $ & $ 0.22 $  & $ 22.64$  \\ 
$ 3  $&$  5/2 $&$   3/2 $&$   2  $&$  5/2  $&$  5/2  $& $ 339.4599960 (70) $ & $ 2.57 $  & $  4.33$  \\ 
$ 3  $&$  5/2 $&$   5/2 $&$   2  $&$  5/2  $&$  3/2  $& $ 339.4626359 (67) $ & $-2.42 $  & $  2.95$  \\ 
$ 3  $&$  5/2 $&$   5/2 $&$   2  $&$  5/2  $&$  5/2  $& $ 339.4759040 (500) $ & $ 0.14 $  & $ 21.24$  \\ 
$ 3  $&$  5/2 $&$   5/2 $&$   2  $&$  5/2  $&$  7/2  $& $ 339.4932119 (70) $ & $ 2.69 $  & $  2.99$  \\ 
$ 3  $&$  5/2 $&$   7/2 $&$   2  $&$  5/2  $&$  5/2  $& $ 339.4992884 (70) $ & $-2.52 $  & $  2.33$  \\ 
$ 3  $&$  5/2 $&$   7/2 $&$   2  $&$  5/2  $&$  7/2  $& $ 339.5166351 (66) $ & $ 0.11 $  & $ 25.35$  \\ 
$ 3  $&$  5/2 $&$   3/2 $&$   2  $&$  3/2  $&$  5/2  $& $ 339.9922571 (56) $ & $-0.33 $  & $  3.89$  \\ 
$ 3  $&$  5/2 $&$   5/2 $&$   2  $&$  3/2  $&$  5/2  $& $ 340.0081263 (38) $ & $-0.69 $  & $ 61.97$  \\ 
$ 3  $&$  5/2 $&$   3/2 $&$   2  $&$  3/2  $&$  3/2  $& $ 340.0196255 (36) $ & $-0.97 $  & $ 92.70$  \\ 
$ 3  $&$  5/2 $&$   7/2 $&$   2  $&$  3/2  $&$  5/2  $& $ 340.0315494 (34) $ & $-0.45 $  & $384.49$  \\ 
$ 3  $&$  5/2 $&$   3/2 $&$   2  $&$  3/2  $&$  1/2  $& $ 340.0354080 (500) $ & $-0.93 $  & $288.68$  \\ 
$ 3  $&$  5/2 $&$   5/2 $&$   2  $&$  3/2  $&$  3/2  $& $ 340.0354080 (500) $ & $-0.62 $  & $323.09$  \\ 
$ 3  $&$  7/2 $&$   7/2 $&$   2  $&$  5/2  $&$  5/2  $& $ 340.2477700 (500) $ & $ 0.45 $  & $379.66$  \\ 
$ 3  $&$  7/2 $&$   9/2 $&$   2  $&$  5/2  $&$  7/2  $& $ 340.2477700 (500) $ & $ 0.31 $  & $413.13$  \\ 
$ 3  $&$  7/2 $&$   5/2 $&$   2  $&$  5/2  $&$  3/2  $& $ 340.2485440 (36) $ & $ 0.62 $  & $367.40$  \\ 
$ 3  $&$  7/2 $&$   5/2 $&$   2  $&$  5/2  $&$  5/2  $& $ 340.2617734 (37) $ & $ 1.01 $  & $ 44.79$  \\ 
$ 3  $&$  7/2 $&$   7/2 $&$   2  $&$  5/2  $&$  7/2  $& $ 340.2649490 (37) $ & $ 0.77 $  & $ 33.50$  \\ 
$ 3  $&$  7/2 $&$   5/2 $&$   2  $&$  5/2  $&$  7/2  $& $ 340.2791201 (54) $ & $ 0.51 $  & $  0.93$  \\ 
\hline \hline

\end{tabular} 
\tablefoot{
\tablefoottext{a}{Line frequencies (with last-digit uncertainties
  between brackets) and Einstein coefficients were
  taken from the fits presented in the CDMS database \citep{Muller01}.}
}
\end{table*}

\section{Results for individual components}
\label{appspecs}

\begin{figure*}
\centering
\includegraphics[width=0.95\textwidth]{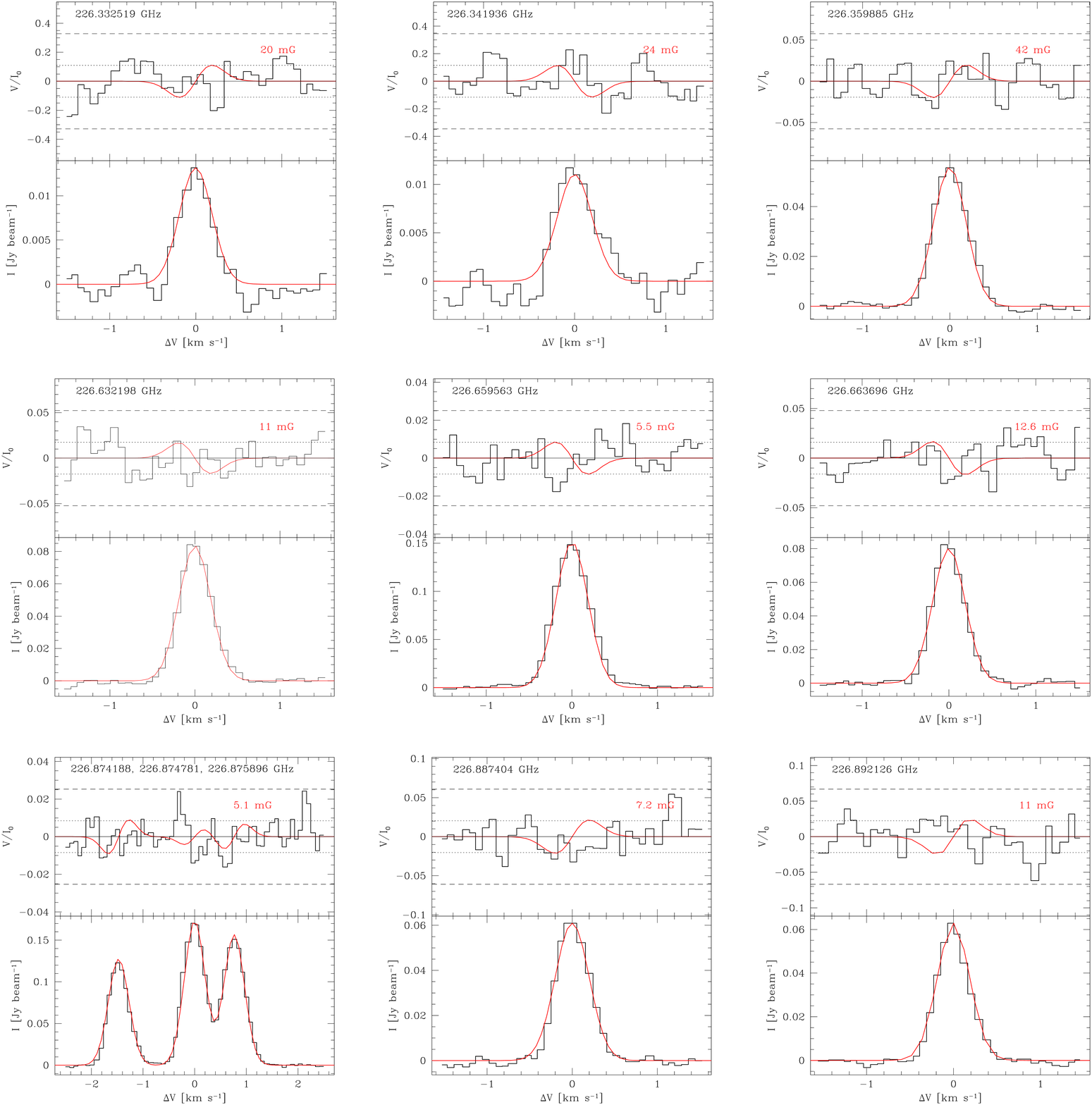}
\caption{Total intensity (I, bottom) and circular polarisation (V,
  top) spectra for all detected CN(2-1) hyperfine components. The
  spectra are extracted at the position of the strongest emission
  peak. The figures are labelled with the component rest frequency as
  determined using line fitting. The $1\sigma$ (short dashed) and
  $3\sigma$ (long dashed) limits are indicated. The $1\sigma$ limit
  magnetic field values are indicated in the figure, and we also
  present the V-spectrum for this field strength in red. Three of the
  components are so close in frequency that they are presented in a single panel.}
\label{applines}
\end{figure*}

Here we present the observed CN $N=2-1$ hyperfine components and their polarisation and magnetic
field limits (in Table~\ref{results}). We also show the total intensity and circular polarisation
spectra (in Fig.~\ref{applines}).

\begin{table*}
\caption{CN components and their polarisation and magnetic field limits}             
\label{results}      
\centering          
\begin{tabular}{l l l c c c c c }     
\hline\hline       
 Transition & Fitted Frequency$^a$ & Catalogue Frequency$^a$ & Offset &
 Peak Flux$^b$ & $m_c$~limit$^c$ & $|B_z|$~limit$^d$ \\
 (CN $N=2-1$) & [GHz] & [GHz] & [MHz] & [mJy beam$^{-1}$] & [$\%$] & [mG ] \\
\hline
$ J=3/2-3/2, F=3/2-5/2$ & $226.3325190 (90)$ & $226.3324986 (56) $
& $+0.020$ & $8.9\pm0.7$ & $11.0$ & $20.0$ \\
$ J=3/2-3/2, F=5/2-3/2$ & $226.3419360 (90)$ & $226.3419298 (54) $
& $+0.006$ & $9.0\pm0.7$ & $11.4$ & $24.0$ \\
$ J=3/2-3/2, F=5/2-5/2$ & $226.3598850 (50)$ & $226.3598710 (500) $
& $+0.014$ & $41.2\pm1.3$ & $1.92$ & $42.0$ \\

$ J=3/2-1/2, F=3/2-3/2$ & $226.6321980 (30)$ & $226.6321901 (35) $
& $+0.008$ & $64.2\pm1.2$ & $1.72$ & $11.0$ \\
$ J=3/2-1/2, F=5/2-3/2$ & $226.6595630 (30)$ & $226.6595584 (26) $
& $+0.005$ & $130.5\pm2.1$ & $0.84$ & $5.5$ \\
$ J=3/2-1/2, F=1/2-1/2$ & $226.6636960 (30)$ & $226.6636928 (25) $
& $+0.003$ & $64.5\pm1.5$ & $1.58$ & $12.6$ \\

$ J=5/2-3/2, F=5/2-3/2$ & $226.8741880 (20)$ & $226.8741908 (23) $
& $-0.001$ & $129.2\pm1.6$ & $1.02$ & $6.7$ \\
$ J=5/2-3/2, F=7/2-5/2$ & $226.8747813 (20)$ & $226.8757813 (30) $
& $[0]$ & $157.6\pm1.6$ & $0.84$ & $11.2$ \\

$ J=3/2-3/2, F=3/2-5/2$ & $226.8758960 (110)$ & $226.8758960 (20) $
& $0.000$ & $101.8\pm7.3$ & $1.30$ & $5.1$ \\
$ J=3/2-3/2, F=3/2-5/2$ & $226.8874040 (40)$ & $226.8874202 (29) $
& $-0.020$ & $47.0\pm1.3$ & $2.06$ & $7.2$ \\
$ J=3/2-3/2, F=3/2-5/2$ & $226.8921260 (40)$ & $226.8921280 (27) $
& $-0.002$ & $46.1\pm1.2$ & $2.39$ & $11.0$ \\
\hline                  
\end{tabular}
\tablefoot{
\tablefoottext{a}{The fitted frequency was derived by a fit to our
  data using the CN $N=2-1, J=5/2-3/2, F=7/2-5/2$ transition as reference, the catalogue frequency is taken from the CDMS catalogue of
  \cite{Muller01}. The uncertainty in the last digits is indicated
  between brackets. The shift between the two is similar to that
  reported for a subset of transitions in \citet{Teague16}.}
\tablefoottext{b}{At the brightest CN emission peak (see text).}
\tablefoottext{c}{The $1\sigma$ limit of the circular polarisation fraction.}
\tablefoottext{d}{The $1\sigma$ magnetic field limit. Because no detection is made,
  we cannot determine the sign of the magnetic field.}}
\end{table*}

\section{Line of node spectra}
\label{lonspec}

\begin{figure*}
   \begin{minipage}[t]{0.5\textwidth}
      \includegraphics[width=\textwidth]{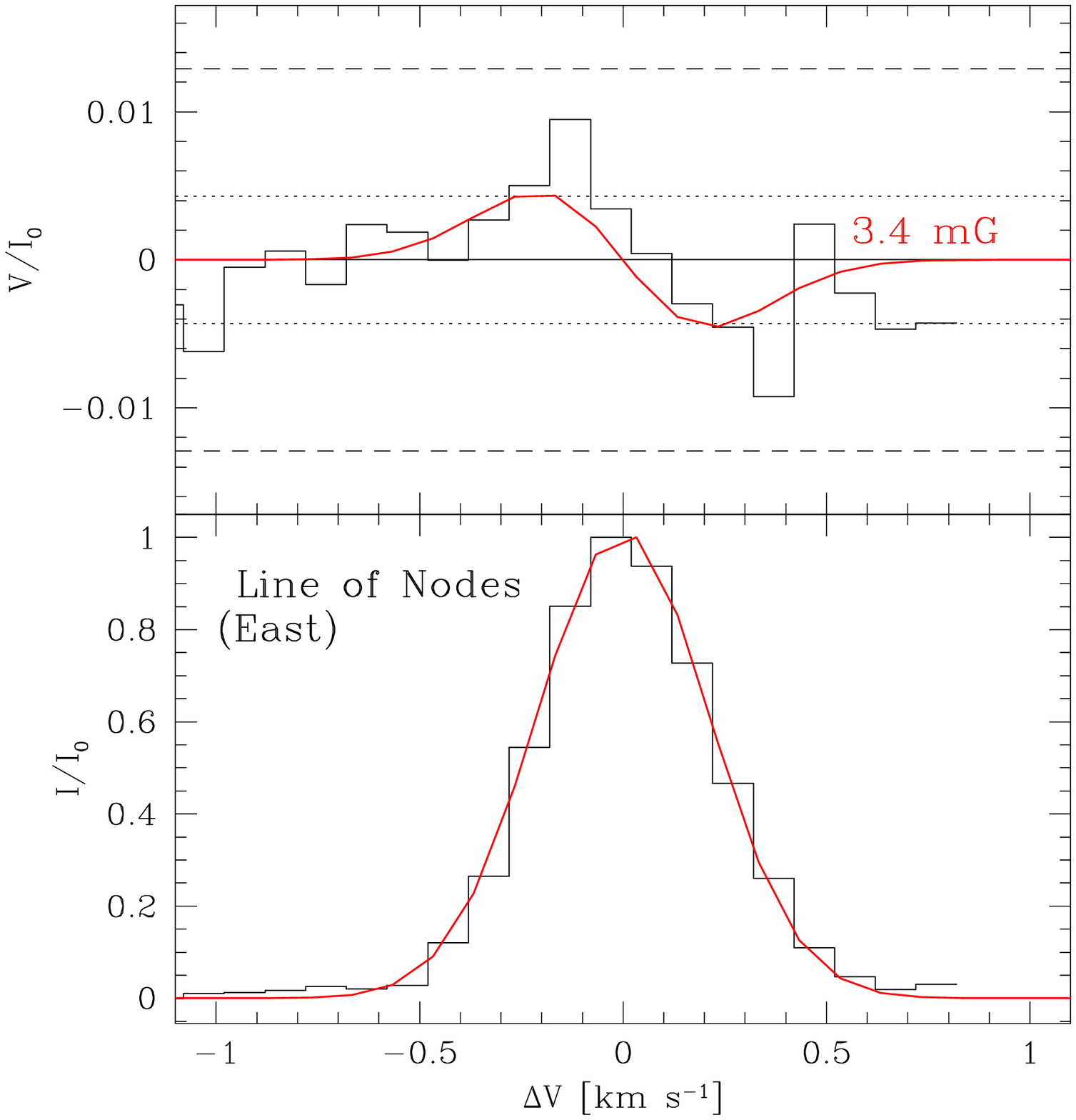}
 \end{minipage}
 \begin{minipage}[t]{0.5\textwidth}
      \includegraphics[width=\textwidth]{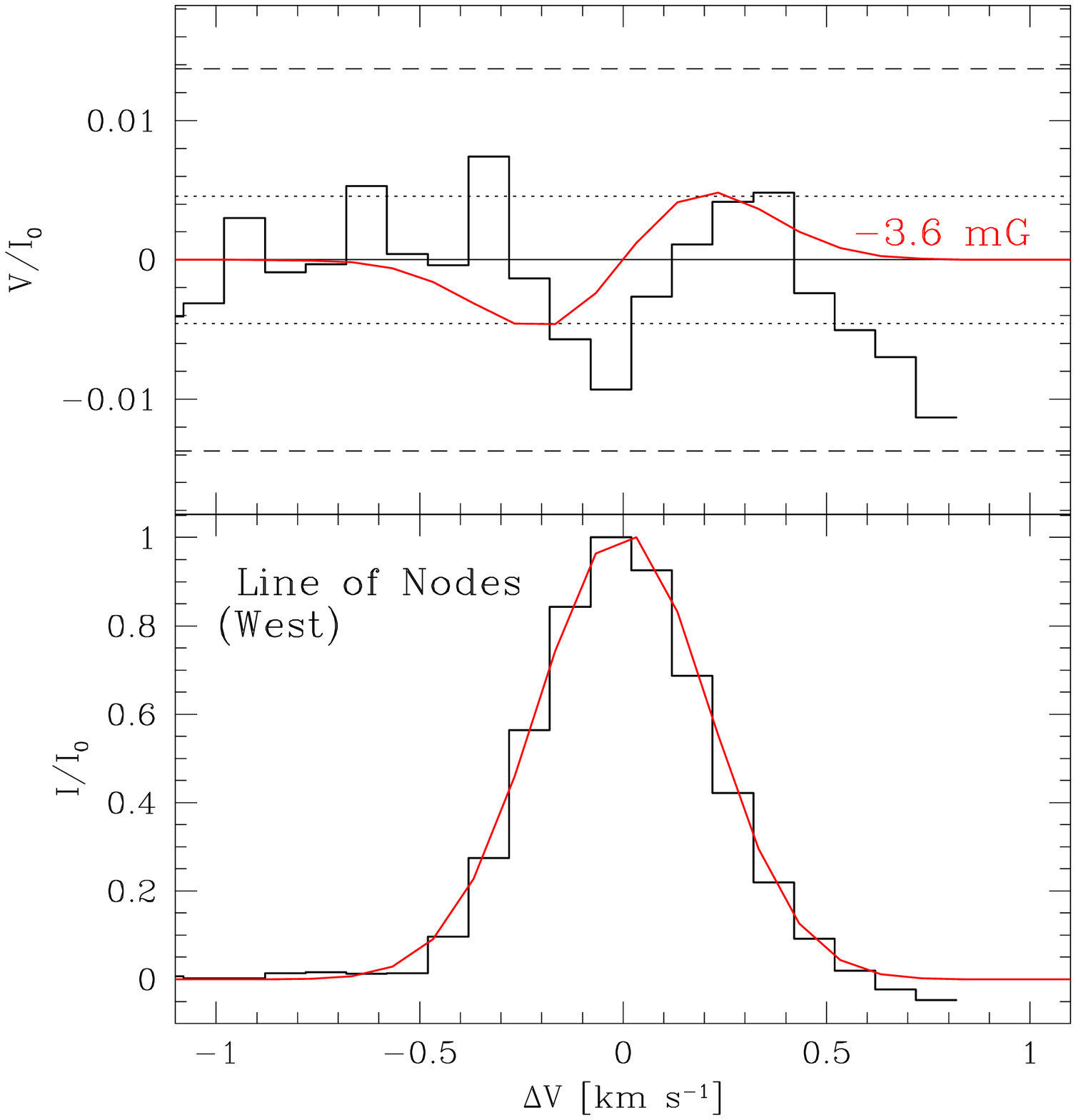}
 \end{minipage}  
\hfill
\caption{Same as the right panel of Fig.~\ref{specs} for emission in the eastern (left) and
 western (right) parts of the line of nodes.}
\label{applon}
\end{figure*}

Here we separately present the stacked spectra of the eastern and western section
of the disc (along the line of nodes; Fig.~\ref{applon}).

\end{appendix}

\end{document}